\documentclass{iau} 
\usepackage{graphicx}
\usepackage{subfigure}
\title[Evolution of the $M_{\bullet} - \sigma$ relation]{Evolution of the $M_{\bullet} - \sigma$ relation}

\author[D. Bhattacharyya \& A. Mangalam]   
{D. Bhattacharyya$^1$
 \and A. Mangalam$^2$}

\affiliation{Indian Institute of Astrophysics\\ Sarjapur Road, Koramangala, Bangalore, 560034, India, \\ email: {\tt $^1$dipanweeta@iiap.res.in, $^2$mangalam@iiap.res.in}} 

\pubyear{2018}
\volume{342}  
\jname{Perseus in Sicily: from black hole to cluster outskirts}
\editors{Keiichi Asada, Elisabete de Gouveia dal Pino, \\ Hiroshi Nagai, Rodrigo Nemmen, Marcello Giroletti, eds.}
\begin{document}

\maketitle
\begin{abstract}
Black holes at the centers of the galaxies grow mainly by the processes of accretion, mergers, and consumption of stars. In the case of gas accretion with cooling sources, the flow is momentum driven, after which the black hole reaches a saturated mass, and subsequently, it grows only by consumption of stars. In addition, we include the effect of mergers on the growth of black hole spin and mass and study its evolution as a function of redshift in a $\Lambda$CDM cosmology using an initial seed mass and spin distribution functions that we have derived. For the stellar ingestion, we have assumed a power-law density profile for the galaxy in our framework of a new relativistic loss cone theory that includes the effect of the black hole spin.  We predict the impact of the evolution on the $M_{\bullet} - \sigma$ relation and compare it with available observations.
\end{abstract}
\vspace{-0.2 in}
\firstsection 
\section{Introduction}
The $M_{\bullet} - \sigma$ relation, $M_{\bullet} \propto \sigma^{p}$, is a very astonishing relation and its origin is still a topic of debate.  \cite[Bhattacharyya \& Mangalam (2018)]{Bhattacharyya_Mangalam2018} [BM18 hereafter] have constructed a static model of deriving the $M_{\bullet} - \sigma $ relation and $M_{\bullet} - M_{b}$ relation together from intensity profiles of 12 elliptical galaxies. Here we discuss the dynamical aspect of the $M_{\bullet} - \sigma$ relation, i. e., the evolution of the relation with redshift and time in $\Lambda$CDM cosmology. From the basic equations of mass and spin evolution of the black hole [\cite[c. f. Mangalam (2015)]{Mangalam_2015}] we have derived the evolution of this relation. For the rate of growth of mass, $\dot{M}_{\bullet}$, we have considered the accretion process, stellar ingestion process and mergers, where, the accretion rate is considered to be a fraction of the Eddington accretion rate, the mass growth by stellar consumption is calculated by using both full and steady loss cone theory [\cite[Mageshwaran \& Mangalam (2015)]{Mageshwaran_Mangalam_2015}] for a galaxy cusp following power law mass density (power law index is $\gamma$) and in case of mergers we have considered both major and minor mergers [\cite[Stewart et al., (2009)]{Stewart_2009}]. We have also considered the BZ torque which contributes to spinning down the black hole while the accretion process spins it up. For spin evolution we have considered only the effect of minor mergers which also contributes in spinning the hole down [\cite[Gammie et al., (2004)]{Gammie_2004}], while we neglect major mergers because the accretion process is the dominant one for spinning up the black hole. We have also incorporated the prescription of saturated mass, $M_{sat}$, given by \cite[King (2003)]{King_2003} which is the critical mass after which the accretion stops and the black hole grows only by stellar ingestion and mergers. We have considered the mergers to be effective from $z \leq 4$ considering the peak of merger activity. Starting from an initial spin, $j_{s}$, an initial mass, $M_{\bullet s}$ we have derived the evolution of the mass and spin of the black hole and its effect on the $M_{\bullet} - \sigma$ relation. 
\begin{figure}
\begin{center}
\subfigure[]{\includegraphics[scale=0.21]{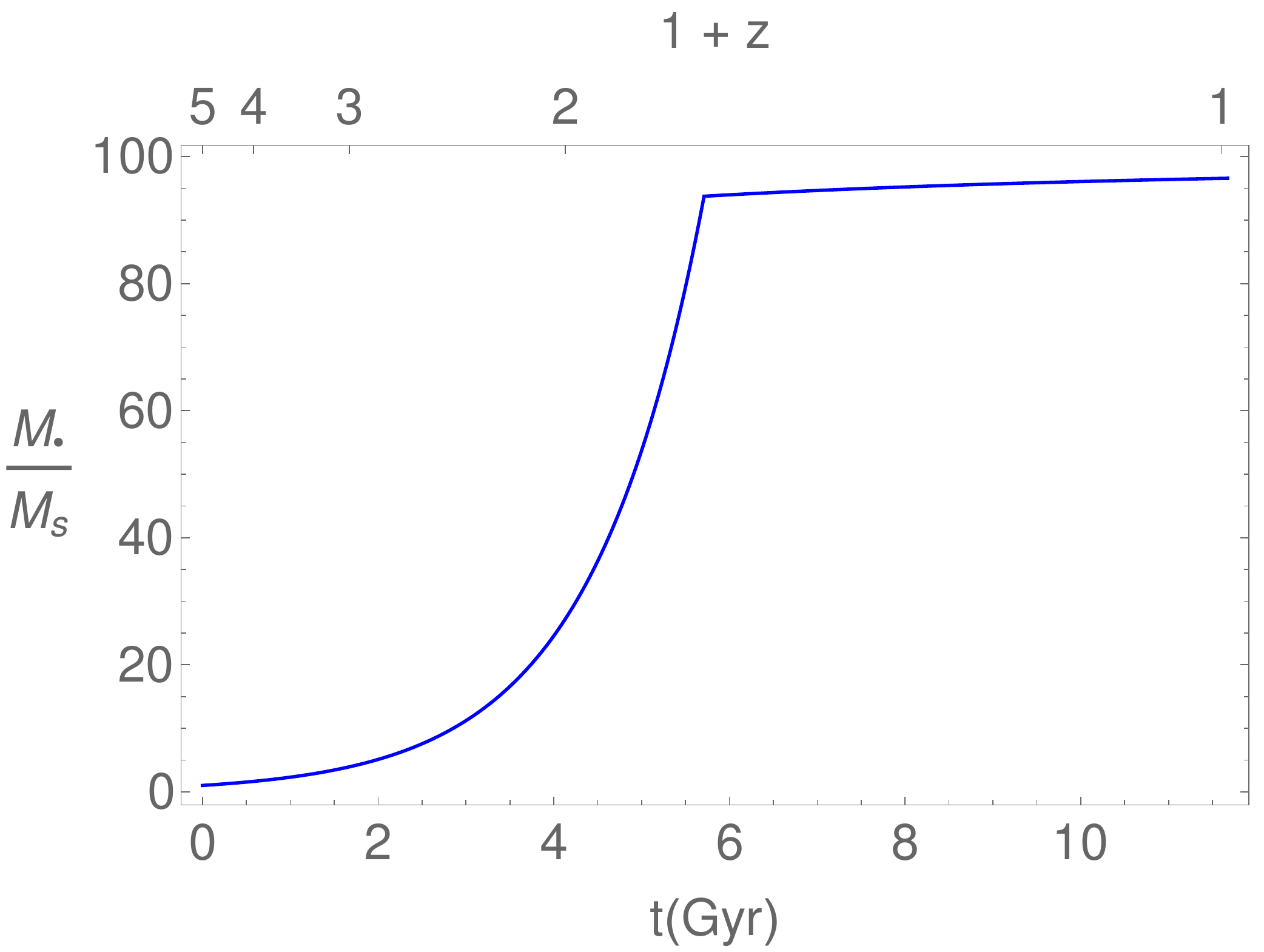}}\label{figa}\hspace{0.15 in}
\subfigure[]{\includegraphics[scale=0.21]{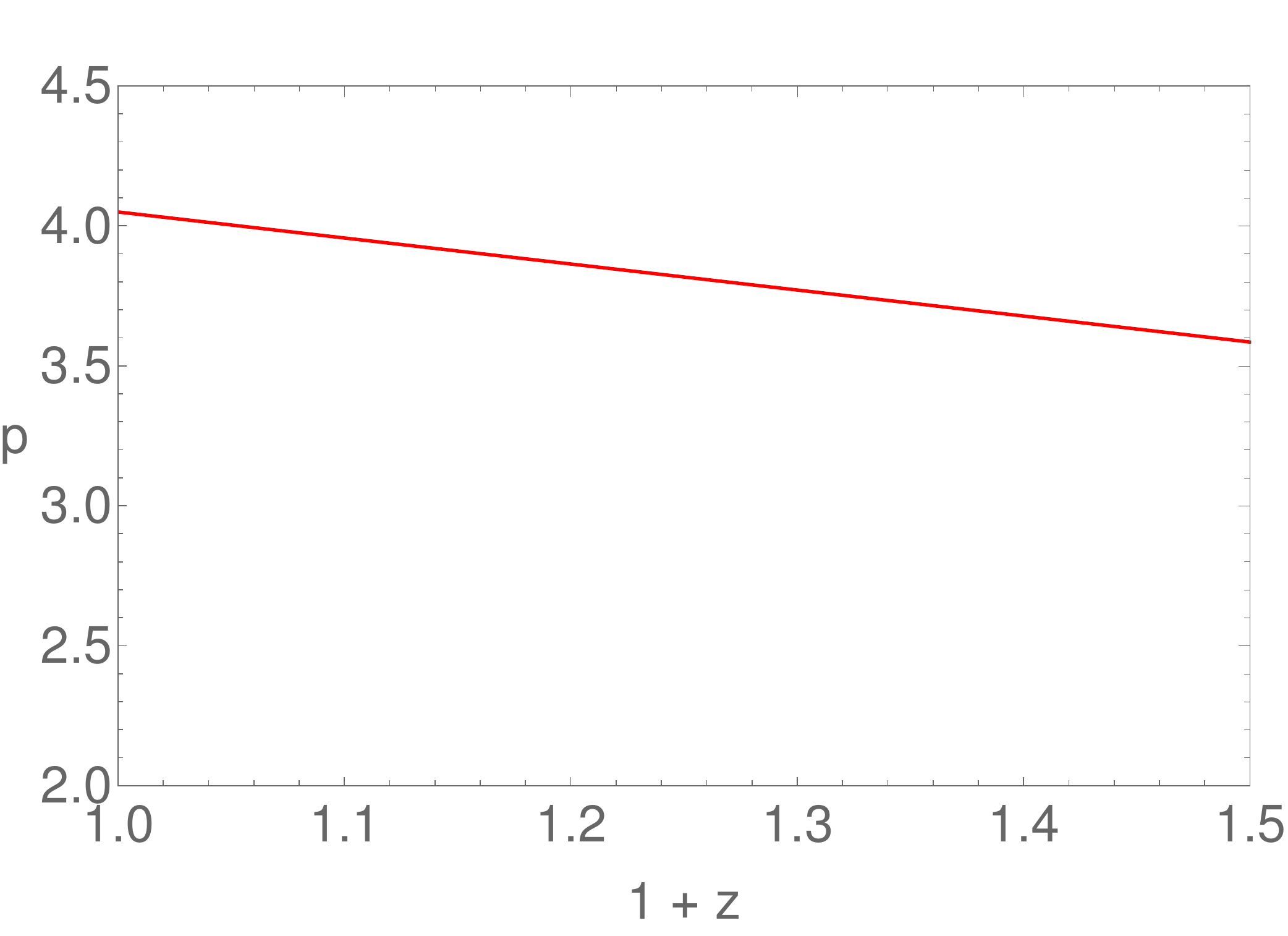}}\label{figb}\vspace{0.1 in}
\subfigure[]{\includegraphics[scale=0.19]{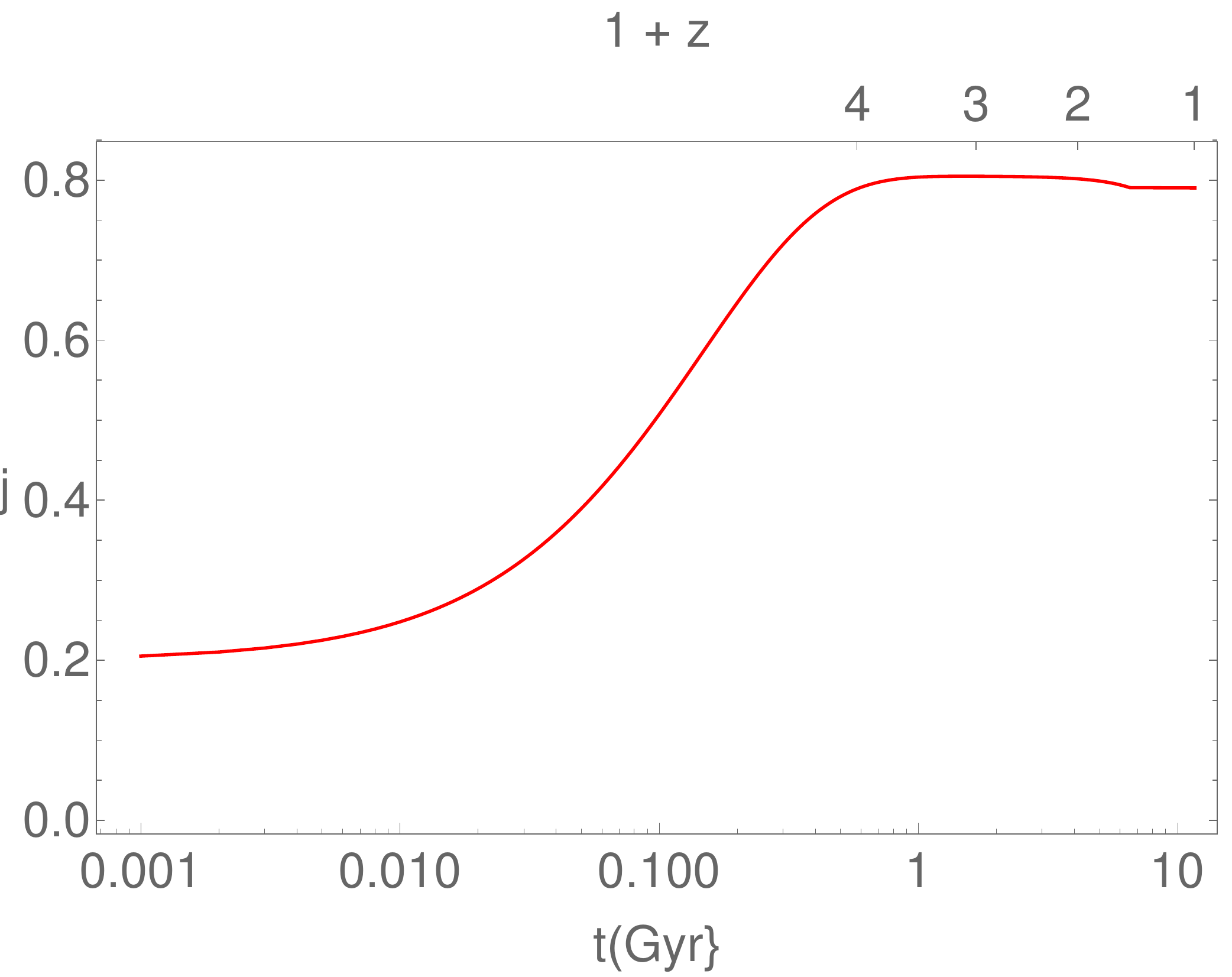}}\label{figc}\hspace{0.15 in}
\subfigure[]{\includegraphics[scale=0.17]{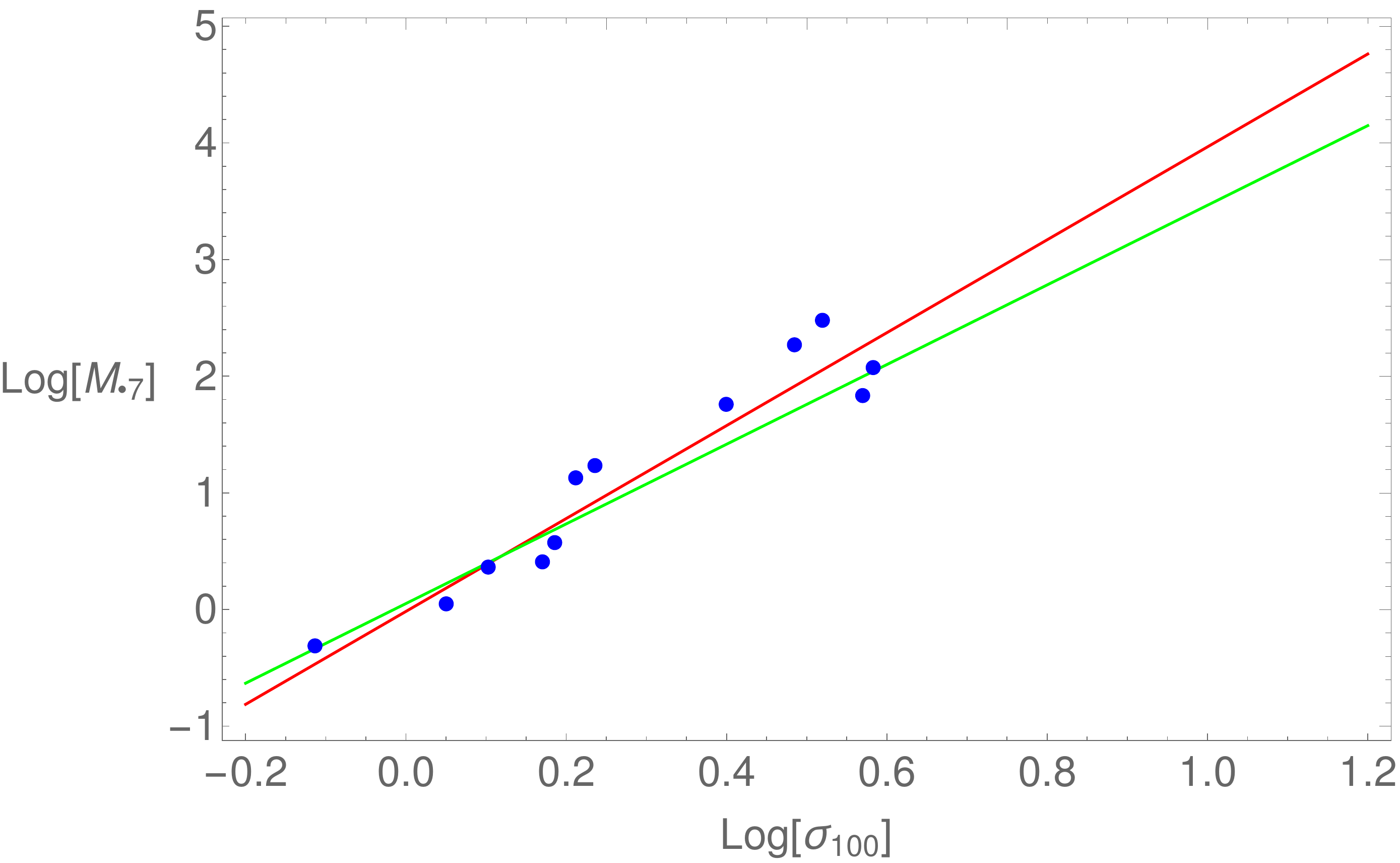}}\label{figd}
\caption{Comic evolution of the mass and the spin of the black hole and the $M_{\bullet} - \sigma$ relation. \textit{(1a, 1c)} Evolution of the mass of the black hole in units of $M_{\bullet s}$ and spin of the black hole as function of time and redshift with merger mass ratio, $q$ = 0.1, $j_{s}$ = 0.2, magnetic field of BZ torque, $B_{4} = B / (10^{4}$ Gauss) = 5, formation redshift, $z_{f}$ = 4 and $\gamma$ = 1.5 and $M_{\bullet s}$ = $10^{5} M_{\odot}$. \textit{(1b)} Evolution of the index $p$ of the $M_{\bullet} \propto \sigma^{p}$ relation. \textit{(1d)} shows $\log(M_{\bullet {7}})$ vs $\log(\sigma_{100})$ for two different redshifts ($z = 0.003$, red and $z = 0.23$, green) along with the data obtained from our calculation in BM18 for 12 elliptical galaxies (their redshift lies in the range 0.004 - 0.002).}
\end{center}
\end{figure}
For a few examples, the resulting mass and spin evolution of the black hole is shown in Figures [\ref{figa}(a), \ref{figc} (c)] and the impact on the index of the $M_{\bullet} - \sigma$ relation which is obtained from the slope of $\log{M_{\bullet}}$ vs $\log{\sigma}$ plot is shown in Figure \ref{figb}(b). We show how our results matches with observations in Figure \ref{figd} (d), where the data of $M_{\bullet}$ and $\sigma$ of 12 elliptical galaxies was obtained from their observed intensity profile in BM18. A more detailed study of the models is discussed in the paper by Bhattacharyya \& Mangalam (2018b) (in preparation).
\vspace{-0.24 in}
\section{Summary and conclusions}
We have obtained the relativistic as well as the non-relativistic evolution of black hole mass and spin as a function of redshift using cosmological $\Lambda CDM$ model for different values of the spin parameter, seed masses and different formation redshifts. For calculation of stellar consumption rate, we use the steady loss cone model. We have compared our model with the available observations of $z$, $M_{\bullet}$ and $\sigma_{||}$ of 12 elliptical galaxies which follow the $M_{\bullet} - \sigma$ relation and we are able to explain the observations from our model. We have computed the evolution of the $M_{\bullet} - \sigma$ relation with redshift by deriving the evolution of the slope and intercept of $\log[M_{\bullet 7}]$ vs $\log[\sigma_{100}]$ plot. Using formula given by \cite[Shankar \etal (2009)]{Shankar_2009}, for renormalization, $p \propto (1 + z)^{-\alpha}$, we show that the $M_{\bullet} - \sigma$ relation holds with $\alpha \simeq $ 0.24 - 0.34 upto $z \simeq 1$ for the index, $p$ lying between 4 - 5. The data from future surveys at high redshift, for example from TMT, can be used to probe the $M_{\bullet} - \sigma$ evolution.


\vspace{-0.24 in}
\begin{thebibliography}{}
\bibitem[Stewart \etal\ (2009)]{Stewart_2009}
{Stewart et al.} 2009,
\textit{ApJ}, 702, 1005

\bibitem[Mangalam (2015)]{Mangalam_2015}
{Mangalam, A., 2015}, 
\textit{ASI conference Series}, 12, 51 - 56



\bibitem[King (2003)]{King_2003}
{King} 2003,
\textit{ApJ}, 596, L27

\bibitem[Bhattacharyya \& Mangalam (2018)]{Bhattacharyya_Mangalam2018}
{Bhattacharyya, D. \&  Mangalam, A.,} 2018, 
\textit{JAA}, 39, Issue 1, article id 4 (2018a)

\bibitem[Gammie \etal\ (2004)]{Gammie_2004}
{Gammie et al.} 2004,
\textit{ApJ}, 602, 312

\bibitem[Mageshwaran \& Mangalam (2015)]{Mageshwaran_Mangalam_2015}
{Mageshwaran, T. \& Mangalam, A., 2015} 2015,
\textit{ApJ}, 814, 141

\bibitem[Shankar \etal (2009)]{Shankar_2009}
{Shankar, F., etal.} 2009,
\textit{ApJ}, 694, 867

\end{thebibliography}
\end{document}